# Robust surface state with unusual protection mechanism in nonsymmorphic semimetal

Xue Liu[1,2], Chunlei Yue[1], Sergey V. Erohin[3,4], Yanglin Zhu[1], Abin Joshy[1], Jinyu Liu[1], Ana M Sanchez[5], David Graf[6], Pavel B. Sorokin[3,4], Zhiqiang Mao[7,1], Jin Hu[8]*, Jiang Wei[1]*

[1]Department of Physics and Engineering Physics, Tulane University, New Orleans, Louisiana 70118, USA

[2]Division of Physics and Applied Physics, School of Physical and Mathematical Sciences, Nanyang Technological University, Singapore 637371, Singapore

[3]National University of Science and Technology "MISIS", Leninsky prospect 4, Moscow 119049, Russian Federation

[4]Moscow Institute of Physics and Technology (State University), 9 Institutskiy per., Dolgoprudny, Moscow Region, 141701, Russian Federation

[5]Department of Physics, University of Warwick, Coventry, CV4 7AL, United Kingdom

[6]National High Magnetic Field Lab, Tallahassee, Florida 32310, USA

[7]Department of Physics, Pennsylvania State University, University Park, Pennsylvania 16802, USA

[8]Department of Physics, Institute for Nanoscience and Engineering, University of Arkansas, Fayetteville, Arkansas 72701, USA

*Corresponding to jinhu@uark.edu; jwei1@tulane.edu,

**In a topological semimetal with Dirac or Weyl points, the bulk-edge correspondence principle predicts a gapless edge mode if the essential symmetry is still preserved at the surface. The detection of such topological surface state has been considered as the fingerprint prove for crystals with nontrivial topological bulk band. On the contrary, it has been proposed that even with symmetry broken at the surface, a new surface band can emerge in nonsymmorphic topological semimetals. The symmetry reduction at the surface lifts the bulk band degeneracies, produces an unusual "floating" surface band with trivial topology. Here, we report quantum transport probing to ZrSiSe thin flakes and reveal transport signatures of this new surface state. Remarkably, though topologically trivial, such a surface band exhibit substantial two-dimensional (2D) Shubnikov–de Haas (SdH) quantum oscillations with high mobility, which signifies a new protection mechanism and may open applications for surface-related devices.**

## I. INTRODUCTION

The surface electronic state has been a central focus of condensed matter physics. Distinct electrical properties from a well-protected surface (or edge) state, such as quantum Hall or quantum spin Hall effects, provide ample opportunities for the surface state-based device applications [1-5]. The conventional surface states, resulting from the termination of the three-dimensional (3D) bulk periodic potential, are susceptible to defects or impurities, which appear nearly inevitably in crystals. Recently, there have been significant breakthroughs in the search for robust surface states along with the search for new topological quantum materials. Topological surface states found in bulk topological insulators [6-11] eliminates backscattering due to the spin-momentum lock, which is originated from the chiral linear energy dispersion as protected by the time-reversal or lattice symmetries. In Weyl semimetals and Dirac semimetals [1, 2], unusual surface states appear as disconnected or jointed Fermi arcs curving in opposite directions, respectively. Experimentally, there has been extensive characterization on the transport properties of surface states in topological insulators [7]. For 3D topological semimetals, Weyl orbit on the surface of bulk Dirac semimetal $Cd_3As_2$ has recently been observed showing quantum oscillations [12, 13] and quantum Hall effect [14, 15] in the nanostructured device owing to the enhanced transport signal ratio of surface to bulk. In contrast to the above topologically protected surface states, a new 2D floating surface state can emerge in ZrSi$M$ ($M$ = S, Se, or Te) nonsymmorphic topological semimetals [4]. Such a new surface state originates from the symmetry reduction at the surface, thus distinct from the well-known "conventional" topological surface state arising from the bulk-edge correspondence principle in topological materials.

ZrSi$M$ belongs to the recently discovered $WHM$-type ($W$ = Zr, Hf, or rare-earth; $H$ = Si, Ge, Sn) [16-24] topological semimetal family. These materials crystallize in layered tetragonal structure (Fig. 1a) and possess two types of Dirac states: the nodal-line Dirac state protected by the $C_{2v}$ symmetry and gaped by spin-orbit coupling [16, 17] and the 2D gapless nodal-point Dirac state protected by the nonsymmorphic symmetry [16, 25]. The different combinations of $W$, $H$, and $M$ elements further give rise to high tunability in spin-orbit coupling [26-28], magnetism [22, 23], and structural dimensionality [17, 21, 27], leading to rich electronic properties of various $WHM$s such as large magnetoresistance [29, 30], high Dirac fermion density [20, 21], strong spin splitting [20], and magnetic field-mediated tunable Dirac and Weyl states [23]. These properties, together with the feasibility in obtaining the atomically thin crystals, make this material family a versatile platform for investigating exotic phenomena of relativistic fermions in nanostructures. In this work, taking advantage of the suppressed bulk contributions in the exfoliated ZrSiSe flakes, we have successfully probed transport of the surface floating band. Unlike the topological nontrivial surface states in many other topological nodal point semimetals, such floating surface band is topologically trivial [4], but exhibit surprisingly quantum oscillations with high mobility which is not generally expected. The robustness of the surface state, as demonstrated both from our transport measurements and density functional theory (DFT) calculations, pave a way for surface-related devices.

## II. RESULTS

### A. Fabrication and characterization of ZrSiSe nano devices

Figure 1a shows the crystal structure of ZrSiSe, which can be viewed as the stacking of Se-Zr-Si-Zr-Se slabs. The week inter-slab bonding strength allows for the mechanical exfoliation of ZrSiSe to atomically thin layers [21], as demonstrated in Fig. 1b. The atomic resolution Scanning Transmission Electron Microscope (STEM) images of the as-exfoliated flakes reveal good crystallinity for the inner parts (Fig. 1c) with shallow amorphous oxidation layers (~5 nm) on the top and bottom surface (Fig. 1d and Supplementary Fig. 1). The stacking of Zr, Si, and Se atoms precisely matches the expected lattice structure of ZrSiSe (Fig. 1c inset).

ZrSiSe devices (Fig. 1b, inset) are fabricated through the standard electron beam lithography. With the magnetic field applied perpendicular to the sample surface (i.e., along the *c*-axis), we observed clear Shubnikov-de Haas (SdH) oscillations in magnetoresistance (MR) for all ZrSiSe thin flakes with various thickness at low temperatures (see Supplementary Fig. 3). Surprisingly, the oscillatory components of the longitudinal resistivity $\Delta\rho_{xx}$, obtained by subtracting background, exhibit different signatures between very thick and thin flakes. In Fig. 2a, we present the $\Delta\rho_{xx}$ for typical thick (176 nm) and thin (28.2 nm) samples. For the thicker sample (176 nm), the oscillation pattern contains only a single frequency of $F_B = 210$ T as revealed by the fast Fourier transform (FFT) analysis (Fig. 2a, inset), which is consistent with the observation in the single crystal bulks [21]. In contrast, the oscillation pattern of the thinner sample (28.2 nm) deviates from the "bulk-like" behavior (Fig. 2a) with an additional frequency occurring around $F_S = 445$ T (Fig. 2a, inset). Although $F_S$ appears to be close to $2 \times F_B$, it should not be regarded as a harmonic frequency of $F_B$ because of their distinct angular dependencies, as will be shown later. Such a surprising, additional frequency component is reproducible for all thin flakes below 60 nm (see Supplementary Fig. 4 and 5).

In principle, a quantum oscillation pattern with a specific frequency corresponds to an extremal Fermi surface cross-section. Therefore, the additional frequency in thin samples indicates that an additional electronic band starts to play a substantial role in transport only in the samples with reduced thickness. The modification of band structure due to 2D quantum confinement is widely observed in 2D materials by reaching the monolayer limit. However, it is unlikely that quantum confinement takes effect at a thickness of ~60 nm, where the $F_S$ component already becomes visible (Fig. 2b). Instead, this unusual frequency is most likely a manifestation of a new surface state owing to a few characteristics. First of all, the signal weight of the $F_S$ band in the transport measurement grows with the decreasing sample thickness. It is shown in Fig. 2b, the relative ratio between the FFT peak amplitudes of $F_S$ and $F_B$ increases significantly when the flake thickness is reduced, indicating the increased weight of the $F_S$ component in thinner samples. This result agrees well with the surface origin of the $F_S$ band and is a natural consequence of the enhanced surface-to-bulk ratio with reducing the thickness, which has also been observed in $Cd_3As_2$ [13].

In addition, the 2D character of the $F_S$ band is in line with a surface state. As shown in Fig. 3a, for a typical sample with a moderate thickness of about 36 nm, the SdH oscillation weakens when the magnetic field is rotated away from the perpendicular direction ($\theta = 0^0$), which is consistent with observations in bulk ZrSiSe [21] and other *WHM* compounds [27, 30, 31]. However, as shown in Figs. 3b, the angular dependences of $F_B$ and $F_S$ obtained from FFT are entirely different: $F_S$ varies significantly with $\theta$, which is distinct from the very weak angular-dependence of $F_B$ and indicates it is not a second-order harmonic of $F_B$. Such angular dependences for $F_B$ and $F_S$ are highly reproducible with various sample thicknesses (Supplementary Fig. 5). To better illustrate the angular dependences of both frequencies in various samples, we have summarized the data in the polar plot shown in Fig. 3c. $F_B$ (blue) appears to be nearly $\theta$-independent up to $\theta = 45°$, consistent with the previous studies on bulk samples [21]. In contrast, $F_S$ at various $\theta$ obtained from different samples are well-aligned to a vertical line (red dashed lines) in the polar plot, i.e., showing a $1/\cos\theta$ dependence. Such a $1/\cos\theta$ dependence implies 2D nature for the $F_S$ band, expected for surface state [13].

Furthermore, the agreement of the oscillation frequency with our DFT calculations and previous ARPES observations [32] provides further support for the surface origin of $F_S$. According to the Onsager relation ($F = \hbar A/2\pi e$), the observed $F_S$ frequency at $\theta = 0°$ corresponds to a Fermi surface cross-section area $A$ of 4.25 $nm^{-2}$, which matches well with the area of the ellipse-like electron pocket around the Brillouin zone X point estimated in our DFT calculations (~4.32 nm$^{-2}$, Fig. 2c), as well as that probed in ARPES experiment (~4.58 nm$^{-2}$, estimated from ref. 32). Given no other Fermi pocket with comparable size can be found either in our DFT calculations or ARPES reports [32], the $F_S$ frequency most likely reflects such an electron pocket of a surface-related state inferred in *WHM* compounds [4].

### B. Mechanism of the surface band generation

Now, we discuss the mechanism of forming such a surface band. Generally, a surface state is expected to be formed as a result of the termination of the bulk potential or surface defects/adsorbates in conventional materials. This possibility can be easily excluded because quantum oscillations, which rely on the formation of complete cyclotron orbits and high mobility (i.e., sharp Landau levels), are generally not expected for "dirty" materials. Given defects or adsorbates are strong scattering centers, quantum oscillation from a surface state is often easily destroyed in conventional materials. However, in ZrSiSe, $F_S$ and its angular dependence in thin flakes are prominent and highly reproducible, even with significant amorphous surface layers observed by STEM (Fig. 1d). Such robustness is inconsistent with the cause of surface chemistry, which often involves some foreign reactant or surface defects.

In addition, in a typical *nodal-point* topological semimetal with isolated bulk Dirac or Weyl points, the bulk-edge correspondence principle results in a gapless mode at the edge when the symmetry group protecting the topology of bulk bands is unbroken on the edge [3]. However, this possibility can also be excluded. ZrSiSe and related *WHM* compounds exhibit the coexistence of nodal-line and nodal-point Dirac states protected by different symmetries [16, 17, 25], but neither of them should lead to a topological surface state. In *WHM* compounds, the nodal-line state is slightly gaped by spin-orbit coupling, and a topological surface state has not been revealed in either DFT calculation [17] or ARPES experiments [16, 18, 32]. Similarly, a topological surface state relevant to the nodal-point Dirac state arising from the bulk-edge correspondence principle is not expected since the corresponding nonsymmorphic symmetry is not preserved at the (001) plane of the crystal [3].

After ruling out the possibility of surface chemistry and bulk-edge correspondence, we argue that this robust $F_S$ surface band revealed in our quantum oscillation experiments represents the recently proposed novel floating surface states derived from the surface symmetry reduction in nonsymmorphic semimetals [4]. Topp *et al.* showed that the ZrSiS bulk symmetry with nonsymmorphic space group *P*4/*nmm* is reduced to the symmorphic wallpaper group *P*4*mm* at the natural cleavage (001) surface. Such nonsymmorphic symmetry reduction significantly deforms the orbital, which lifts the degeneracy of the bulk bands at Brillouin zone X point and consequently causes an unpinned surface band floating on top of the bulk band [4]. Such a proposed floating surface state is quantitatively consistent with the ARPES observations of the Fermi pocket with the 2D character at X [4, 16]. The isostructural compound ZrSiSe studied in this work also exhibits electron pocket at X point with similar surface states, as revealed by our DFT calculations (see Supplementary Note 1).

### C. Properties of the surface floating band

The properties of the $F_S$ band provide further support for this argument. The surface floating band is formed by lifting the degeneracy of the bulk band and is thus topological trivial [4], which can be revealed by the Berry phase analysis. We have separated the $F_B$ and $F_S$ oscillation components and extracted the Berry phase for both bands using the Landau fan diagram (see Methods). As shown in Fig. 4a, for ZrSiSe flakes with a range of thickness, the linear fits of the Landau indices *n* yield intercepts $n_0$ around 0 and -0.5 for $F_S$ and $F_B$ bands, respectively. Berry phase $\phi_B$ can be derived via $2\pi(n_0+\delta)$, where $\delta = \pm 1/8$ for the 3D band (e.g., the bulk $F_B$ band) and 0 for the 2D band (e.g., the surface $F_S$ band). As summarized in Fig. 4b, the Berry phase is trivial ($\phi_B^{surface} \approx 0$) for the surface $F_S$ band in each sample, in sharp contrast with that of the bulk $F_B$ band which exhibits an average Berry phase of $\phi_B^{Bulk} \approx -0.68\pi \pm 1/4\ \pi$. This result is further verified through directly fitting the oscillation pattern using the multiband Lifshitz-Kosevich model (see Methods), which confirms the distinct topology of the bulk and the surface floating bands in ZrSiSe.

Furthermore, the effective cyclotron mass of the $F_S$ band also agrees with the scenario of the surface floating band. The formation of the floating band at the surface of our material can be modeled by breaking the nonsymmorphic glide plane symmetry and introducing a large mass for S and Zr orbitals [4], so such surface state is expected to be more massive, which is indeed observed in ZrSiSe. We have extracted effective masses for both bulk $F_B$ and surface $F_S$ bands from the temperature dependence of FFT amplitude for ZrSiSe samples with various thicknesses (see Methods) (Fig. 4c). As summarized in Fig. 4d, the effective cyclotron mass for the surface floating band $m_S^*$ is around $0.39 m_0$ ($m_0$ denotes free electron mass) for all analyzed samples, which is around twice as large as that of the bulk band ($m_B^* \sim 0.19\ m_0$).

### D. Roubustness discussions of the surface floating band

The above discussions have established that the additional $F_S$ component observed in the quantum oscillation of ZrSiSe nano-flakes originates from the surface floating band. The observation of quantum oscillations caused by such topological trivial surface state is unusual because the lack of a protection mechanism is generally expected to lead to a vulnerable surface state with low mobility that is not favorable to quantum oscillations. Despite of the apparent surface degradations (Fig. 1d and Supplementary Figure 1), the LK-fitting (see Methods) has revealed high quantum mobility of $1.20 \times 10^3$ cm$^2$V$^{-1}$s$^{-1}$ at 1.7K for the topologically trivial surface $F_S$ band, which is comparable with the topologically protected bulk band ($1.74 \times 10^3$ cm$^2$V$^{-1}$s$^{-1}$). The high quantum mobility for the surface band is consistent with the transport mobility of $1.84 \times 10^3$ cm$^2$V$^{-1}$s$^{-1}$ estimated from the multichannel model of Hall effect (Fig. 4e and f) (see Methods). This result implies minimized surface scattering of charge carriers caused by surface deformation or disorders. Indeed, it is consistent with our STEM observations in Fig. 1d, which shows an atomically sharp interface between the oxidized amorphous layer and the inner crystalline layer. On the other hand, the immunity from the surface disorder is further invested by DFT calculations. As discussed in Supplementary Note 2, the surface band is still preserved even by replacing all the Se atoms in the topmost surface with [OH] or [O$_4$]. Such robustness of the surface band itself is unexpected and may benefit from a sort of protection mechanism that deserves further investigations. One possible interpretation could be the connection with the bulk topological band: given the surface floating band for nonsymmorphic ZrSiSe is caused by lifting degeneracy of the bulk band at the surface, it could be robust when the corresponding bulk band is topologically protected. In ZrSiSe, the surface floating band is related to the bulk Dirac band protected by the nonsymmorphic symmetry. Therefore, an "indirectly" protected surface bands with trivial topology could appear in ZrSiSe, which represents a novel protection mechanism in crystalline solid with similar nonsymmorphic symmetry.

### III. SUMMARY

In summary, we have systematically studied quantum oscillations of exfoliated ZrSiSe nano-flakes and successfully detected a new 2D, trivial surface state, which can be attributed to the surface floating state caused by symmetry reduction at the surface. Our results also suggest such a surface is trivial but robust and likely protected via a new mechanism. Our findings provide a new arena for the study of exotic surface states in topological quantum materials, which is an important step towards practical application in modern electronics and surface-related devices.

### IV. METHODS

#### A. Sample preparation

The ZrSiSe single crystal was synthesized by using a chemical vapor transport (CVT) method. The stoichiometric mixture of Zr, Si and Se powder was sealed in a quartz tube with iodine being used as a transport agent (2 mg/cm$^3$). Plate-like single crystals with metallic luster can be obtained via the vapor transport growth with a temperature gradient from 950 ºC to 850 ºC. The composition and phase of the single crystals were examined by Energy-dispersive x-ray spectroscopy and x-ray diffraction, respectively. The thin flakes of ZrSiSe were obtained through a micromechanical exfoliation. The thickness of thin flakes was precisely determined by an atomic force microscope. The ZrSiSe devices with the standard four-terminal resistivity or six-terminal Hall bar geometry were fabricated by using the standard electron beam lithography, followed by the deposition of 5nm Ti/ 50nm Au as contacts via electron beam evaporation. Ohmic contacts of devices were achieved by current annealing before the transport measurements.

### B. Scanning Transmission Electron Microscopy

Atomic-resolution Annular Dark Field STEM images of the flakes were recorded with a JEOL ARM200F over collection angles 45–180 mrad. High signal-to-noise images were formed by averaging multiple, rapidly acquired frames to remove scan distortions.

### C. Current annealing process

The fabricated ZrSiSe devices were conducted current annealing between every two electrodes of the Hall bar device geometry, which has been used for cleaning or improving contacts of graphene devices [33, 34]. By using a parameter analyzer, a DC voltage was applied and swept step by step from 0V up to 2.5V with the current recorded. A typical annealing process has been shown by Supplementary Fig.2, before annealing the I-V sweeps show nonlinear behavior with high resistance ~0.4 MΩ indicating a large contact resistance caused by the oxidation layer between the metal and ZrSiSe single crystals; when the voltage sweeps up around 2.4V, the current suddenly jumps to 1 mA which is the preset current limitation. Then, after stabilizing for a few sweeps, the I-V sweep becomes linear, and the two-probe resistance decreases down to ~1 KΩ. By processing current annealing, the oxidation layer under the electrodes has been broken down to achieve nearly ohmic contact between metal and ZrSiSe crystals.

### D. Magnetotransport measurements

Before the high field experiments, the ZrSiSe devices were tested by an in house 9T-PPMS. The high field magnetotransport measurements were performed at National High Magnetic Field Laboratory (NHMFL) in Tallahassee by using an 18T superconducting magnet and a 31T resistive magnet. The AC current used for all devices was between 20 μA to 50 μA supplied by Keithley 6221 AC and DC Current Source. The longitudinal/transverse voltages were measured using lock-in amplifiers with the frequencies triggered by the AC currents. The noise ratio was reduced by twisted pairs between two voltage cables and two current cables, respectively.

### E. Landau level fan diagrams

To examine the Berry phase $\phi_B$ accumulated along cyclotron orbits for bulk and surface bands, we performed Landau Level fan diagram analysis using the longitudinal conductivity $\sigma_{xx}$, which was derived via $\sigma_{xx} = \rho_{xx}/(\rho_{xx}^2 + \rho_{xy}^2)$ where $\rho_{xx}$ and $\rho_{xy}$ are longitudinal and transverse resistivity, respectively, as shown in Supplementary Fig 6. Bulk and surface band's oscillations are then separated by FFT filter, in order to build the LL fan diagram for each band. Given the integer LL indices should be assigned when the Fermi level lies between two adjacent LLs where the density of state reacthes a minimum [35, 36], we assign the integer Landau indices to the oscillation maximum of $\sigma_{xx}$ [37], as shown in Supplementary Fig 7. Berry phase for each band was extracted from the intercept of the linear fit of the LL fan diagram. The slope of each linear fit yields the oscillation frequency, which only differs from the frequency obtained from the FFT by 1-2%. Such consistency indicates the reliability of the obtained intercept and the derived Berry phase [36].

### F. Lifshitz-Kosevich fit

In addition to the LL index fan diagram, we also performed the direct Lifshitz-Kosevich (LK) fitting to extract the Berry phase. In our ZrSiSe system, the SdH oscillations are treated as the linear superposition of bulk and surface frequency ($F_B$, $F_S$) oscillations. Each frequency component can be described as [37-39]:

$$\Delta\rho_{xx} = A\frac{5}{2}(\frac{B}{2F})^{\frac{1}{2}} e^{-\lambda_D} \frac{\lambda}{\sinh\lambda} \cos[2\pi(\frac{F}{B} + \gamma - \delta)]$$

The thermal damping factor $\lambda = 2\pi^2 k_B T m^*/(\hbar eB)$, $\lambda_D = 2\pi^2 k_B T_D m^*/(\hbar eB)$, where $m^*$ is the effective mass, $T_D$ is Dingle temperature, $F$ is the frequency, $\gamma$ is defined as $\gamma = 0.5 - \phi_B/2\pi$, and $\delta$ is the phase shift, which is $\pm 1/8$ for 3D and 0 for 2D. Typical fittings for 36 nm and 46.2 nm samples are shown in Supplementary Fig. 8. The extracted Dingle temperature for bulk and surface states are 6.49 K and 4.57 K, respectively, from which the quantum relaxation time $\tau_q = \hbar/(2\pi k_B T_D)$ and quantum mobility $\mu_q = e\tau_q/m^*$ can be extracted. Moreover, the fitting also yields Berry phases of -1.09π (δ = -1/8) or -0.59π ($\delta_{3D}$ = 1/8) for the bulk and 0.1π ($\delta_{2D}$ = 0) for the surface. These values are in good agreement with those derived from the LL fan diagram.

### G. Effective mass

The effective masses of bulk and surface bands for various samples were obtained from the temperature dependence of the quantum oscillations (1.7 to 20 K), as shown in Supplementary Fig. 9, by fitting the FFT peak intensity to the thermal damping term of the LK-formula [37].

### H. Transport mobility from Hall effect

As illustrated in Fig. 4e, the Hall signal is contributed from bulk electron, bulk hole, and surface electron (as confirmed by ARPES measurements [34]). Then, the conductivity and hall coefficient can be described by a three-channel model [40]:

$$\sigma_{xx} = en_p\mu_p - en_n\mu_n - en_t\mu_t,$$

$$R_H = \frac{n_p\mu_p^2 - n_n\mu_n^2 - n_t\mu_t^2}{e[n_p\mu_p + n_n\mu_n + n_t\mu_t]^2}$$

with $n_n = n_e \times d$, $n_p = n_h \times d$. Here, $n_e$, $n_h$ are the hole and electron carrier density of bulk state, respectively, and d is the sample thickness. $n_t$ is the surface state carrier density. $\mu_n$, $\mu_p$, $\mu_t$ are bulk hole, bulk electron, and surface electron mobilities, respectively. By fitting the thickness dependent $\sigma_{xx}$ and $R_H$ (as plotted in Fig. 4f), we can extract $\mu_t$=1.84×10$^3$ cm$^2$V$^{-1}$s$^{-1}$ for the surface band and $\mu_p$=1.39×10$^4$ cm$^2$V$^{-1}$s$^{-1}$, $\mu_n$=1.03×10$^4$ cm$^2$V$^{-1}$s$^{-1}$ for bulk band. These derived quantum and transport mobilities of the surface electrons are comparable with that of the topological protected bulk electrons in many other topological semimetals [37].

# REFERENCES

[1]. Yan, B. & Felser, C. Topological Materials: Weyl Semimetals. *Annual Review of Condensed Matter Physics* **2017**, 8, 337-354.

[2]. Armitage, N. P., Mele, E. J. & Vishwanath, A. Weyl and Dirac semimetals in three-dimensional solids. *Rev. Mod. Phys.* **2018**, 90, 015001.

[3]. Fang, C., Weng, H., Dai, X. & Fang, Z. Topological nodal line semimetals. *Chin. Phys. B* **2016**, 25, 117106.

[4]. Topp, A., Queiroz, R., Grüneis, A., Müchler, L., Rost, A. W., Varykhalov, A., Marchenko, D., Krivenkov, M., Rodolakis, F., McChesney, J. L., Lotsch, B. V., Schoop, L. M. & Ast, C. R. Surface Floating 2D Bands in Layered Nonsymmorphic Semimetals: ZrSiS and Related Compounds. *Phys. Rev. X* **2017**, 7, 041073.

[5]. Kou, X., Fan, Y. & Wang, K. L. Review of Quantum Hall Trio. *J. Phys. Chem. Solids* **2017**.

[6]. Hasan, M. Z. & Kane, C. L. Colloquium: Topological insulators. *Rev. Mod. Phys.* **2010**, 82, 3045-3067.

[7]. Qi, X.-L. & Zhang, S.-C. Topological insulators and superconductors. *Rev. Mod. Phys.* **2011**, 83, 1057-1110.

[8]. Xu, S. Y., Liu, C., Kushwaha, S. K., Sankar, R., Krizan, J. W., Belopolski, I., Neupane, M., Bian, G., Alidoust, N., Chang, T. R., Jeng, H. T., Huang, C. Y., Tsai, W. F., Lin, H., Shibayev, P. P., Chou, F. C., Cava, R. J. & Hasan, M. Z. Observation of Fermi arc surface states in a topological metal. *Science* **2015**, 347, 294-298.

[9]. Lv, B. Q., Weng, H. M., Fu, B. B., Wang, X. P., Miao, H., Ma, J., Richard, P., Huang, X. C., Zhao, L. X., Chen, G. F., Fang, Z., Dai, X., Qian, T. & Ding, H. Experimental Discovery of Weyl Semimetal TaAs. *Phys. Rev. X* **2015**, 5, 031013.

[10]. Lv, B. Q., Xu, N., Weng, H. M., Ma, J. Z., Richard, P., Huang, X. C., Zhao, L. X., Chen, G. F., Matt, C. E., Bisti, F., Strocov, V. N., Mesot, J., Fang, Z., Dai, X., Qian, T., Shi, M. & Ding, H. Observation of Weyl nodes in TaAs. *Nature Phys.* **2015**, 11, 724–727.

[11]. Xu, S., Belopolski, I., Alidoust, N., Neupane, M., Bian, G., Zhang, C., Sankar, R., Chang, G., Yuan, Z., Lee, C., Huang, S., Zheng, H., Ma, J., Sanchez, D. S., Wang, B., Bansil, A., Chou, F., Shibayev, P. P., Lin, H., Jia, S. & Hasan, M. Z. Discovery of a Weyl fermion semimetal and topological Fermi arcs. *Science* **2015**, 7, 613-617.

[12]. Potter, A. C., Kimchi, I. & Vishwanath, A. Quantum oscillations from surface Fermi arcs in Weyl and Dirac semimetals. *Nature Commun.* **2014**, 5, 5161.

[13]. Moll, P. J. W., Nair, N. L., Helm, T., Potter, A. C., Kimchi, I., Vishwanath, A. & Analytis, J. G. Transport evidence for Fermi-arc-mediated chirality transfer in the Dirac semimetal $Cd_3As_2$. *Nature* **2016**, 535, 266-270.

[14]. Zhang, C., Zhang, Y., Yuan, X., Lu, S., Zhang, J., Narayan, A., Liu, Y., Zhang, H., Ni, Z., Liu, R., Choi, E. S., Suslov, A., Sanvito, S., Pi, L., Lu, H.-Z., Potter, A. C. & Xiu, F. Quantum Hall effect based on Weyl orbits in $Cd_3As_2$. *Nature* **2019**, 565, 331-336.

[15]. Wang, C. M., Sun, H. P., Lu, H. & Xie, X. C. 3D Quantum Hall Effect of Fermi Arcs in Topological Semimetals. *Physical Review Letters* **2017**, 119, 136806.

[16]. Schoop, L. M., Ali, M. N., Straszer, C., Topp, A., Varykhalov, A., Marchenko, D., Duppel, V., Parkin, S. S. P., Lotsch, B. V. & Ast, C. R. Dirac cone protected by non-symmorphic symmetry and three-dimensional Dirac line node in ZrSiS. *Nature Commun.* **2016**, 7, 11696.

[17]. Xu, Q., Song, Z., Nie, S., Weng, H., Fang, Z. & Dai, X. Two-dimensional oxide topological insulator with iron-pnictide superconductor LiFeAs structure. *Phys. Rev. B* **2015**, 92, 205310.

[18]. Neupane, M., Belopolski, I., Hosen, M. M., Sanchez, D. S., Sankar, R., Szlawska, M., Xu, S.-Y., Dimitri, K., Dhakal, N., Maldonado, P., Oppeneer, P. M., Kaczorowski, D., Chou, F., Hasan, M. Z. & Durakiewicz, T. Observation of Topological Nodal Fermion Semimetal Phase in ZrSiS. *Phys. Rev. B* **2016**, 93, 201104.

[19]. Takane, D., Wang, Z., Souma, S., Nakayama, K., Trang, C. X., Sato, T., Takahashi, T. & Ando, Y. Dirac-node arc in the topological line-node semimetal HfSiS. *Phys. Rev. B* **2016**, 94, 121108.

[20]. Hu, J., Tang, Z., Liu, J., Zhu, Y., Wei, J. & Mao, Z. Nearly massless Dirac fermions and strong Zeeman splitting in the nodal-line semimetal ZrSiS probed by de Haas-van Alphen quantum oscillations. *Physical Review B* **2017**, 96, 045127.

[21]. Hu, J., Tang, Z., Liu, J., Liu, X., Zhu, Y., Graf, D., Myhro, K., Tran, S., Lau, C. N., Wei, J. & Mao, Z. Evidence of Topological Nodal-Line Fermions in ZrSiSe and ZrSiTe. *Physical Review Letters* **2016**, 117, 016602.

[22]. Hosen, M. M., Dhakal, G., Dimitri, K., Maldonado, P., Aperis, A., Kabir, F., Sims, C., Riseborough, P., Oppeneer, P. M., Kaczorowski, D., Durakiewicz, T. & Neupane, M. Discovery of topological nodal-line fermionic phase in a magnetic material GdSbTe. *Sci. Rep.* **2018**, 8, 13283.

[23]. Schoop, L. M., Topp, A., Lippmann, J., Orlandi, F., Müchler, L., Vergniory, M. G., Sun, Y., Rost, A. W., Duppel, V., Krivenkov, M., Sheoran, S., Manuel, P., Varykhalov, A., Yan, B., Kremer, R. K., Ast, C. R. & Lotsch, B. V. Tunable Weyl and Dirac states in the nonsymmorphic compound CeSbTe. *Sci. Adv.* **2018**, 4, eaar2317.

[24]. Hosen, M. M., Dimitri, K., Aperis, A., Maldonado, P., Belopolski, I., Dhakal, G., Kabir, F., Sims, C., Hasan, M. Z., Kaczorowski, D., Durakiewicz, T., Oppeneer, P. M. & Neupane, M. Observation of gapless Dirac surface states in ZrGeTe. *Phys. Rev. B* **2018**, 97, 121103.

[25]. Topp, A., Lippmann, J. M., Varykhalov, A., Duppel, V., Lotsch, B. V., Ast, C. R. & Schoop, L. M. Non-symmorphic band degeneracy at the Fermi level in ZrSiTe. *New J. Phys.* **2016**, 18, 125014.

[26]. Chen, C., Xu, X., Jiang, J., Wu, S. C., Qi, Y. P., Yang, L. X., Wang, M. X., Sun, Y., Schröter, N. B. M., Yang, H. F., Schoop, L. M., Lv, Y. Y., Zhou, J., Chen, Y. B., Yao, S. H., Lu, M. H., Chen, Y. F., Felser, C., Yan, B. H., Liu, Z. K. & Chen, Y. L. Dirac line nodes and effect of spin-orbit coupling in the nonsymmorphic critical semimetals *M*SiS (*M* = Hf, Zr). *Phys. Rev. B* **2017**, 95, 125126.

[27]. Hu, J., Zhu, Y. L., Graf, D., Tang, Z. J., Liu, J. Y. & Mao, Z. Q. Quantum oscillation studies of topological semimetal candidate ZrGe*M* (*M* = S, Se, Te). *Phys. Rev. B* **2017**, 95, 205134.

[28]. Topp, A., Vergniory, M. G., Krivenkov, M., Varykhalov, A., Rodolakis, F., McChesney, J. L., Lotsch, B. V., Ast, C. R. & Schoop, L. M. The effect of spin-orbit coupling on nonsymmorphic square-net compounds. *J. Phys. Chem. Solids* **2017**.

[29]. Singha, R., Pariari, A., Satpati, B. & Mandal, P. Large nonsaturating magnetoresistance and signature of nondegenerate Dirac nodes in ZrSiS. *Proc. Natl. Acad. Sci. USA* **2017**, 114, 2468-2473.

[30]. Ali, M. N., Schoop, L. M., Garg, C., Lippmann, J. M., Lara, E., Lotsch, B. & Parkin, S. Butterfly Magnetoresistance, Quasi-2D Dirac Fermi Surfaces, and a Topological Phase Transition in ZrSiS. *Sci. Adv.* **2016**, 2, e1601742.

[31]. Kumar, N., Manna, K., Qi, Y., Wu, S.-C., Wang, L., Yan, B., Felser, C. & Shekhar, C. Unusual magnetotransport from Si-square nets in topological semimetal HfSiS. *Phys. Rev. B* **2017**, 95, 121109(R).

[32]. Hosen, M. M., Dimitri, K., Belopolski, I., Maldonado, P., Sankar, R., Dhakal, N., Dhakal, G., Cole, T., Oppeneer, P. M., Kaczorowski, D., Chou, F., Hasan, M. Z., Durakiewicz, T. & Neupane, M. Tunability of the topological nodal-line semimetal phase in ZrSiX-type materials (X=S, Se, Te). *Phys. Rev. B* **2017**, 95, 161101.

[33]. Moser, J., Barreiro, A. & Bachtold, A. Current-induced cleaning of graphene. *Appl Phys Lett* **2007**, 91.

[34]. Ramamoorthy, H. & Somphonsane, R. In-situ current annealing of graphene-metal contacts. *J. Phys.: Conf. Ser.* **2018**, 1144, 012186.

[35]. Xiong, J., Luo, Y. K., Khoo, Y. H., Jia, S., Cava, R. J. & Ong, N. P. High-field Shubnikov-de Haas oscillations in the topological insulator Bi2Te2Se. *Physical Review B* **2012**, 86.

[36]. Ando, Y. Topological Insulator Materials. *J. Phys. Soc. Jpn.* **2013**, 82, 102001.

[37]. Hu, J., Xu, S., Ni, N. & Mao, Z. Electronic Transport and quantum oscillation of Topological Semimetals. *Annu. Rev. Mater. Res.*, **2019**, 49, 207-252.

[38]. Shoenberg, D., (Cambridge Univ. Press, 1984).
[39]. Murakawa, H., Bahramy, M. S., Tokunaga, M., Kohama, Y., Bell, C., Kaneko, Y., Nagaosa, N., Hwang, H. Y. & Tokura, Y. Detection of Berry's Phase in a Bulk Rashba Semiconductor. *Science* **2013**, 342, 1490-1493.
[40]. Backes, D., Huang, D. H., Mansell, R., Lanius, M., Kampmeier, J., Ritchie, D., Mussler, G., Gumbs, G., Grutzmacher, D. & Narayan, V. Disentangling surface and bulk transport in topological-insulator p-n junctions. *Phys Rev B* **2017**, 96.

**ACKNOWLEDGEMENT**

The work at Tulane is supported by the US Department of Energy under grant DE-SC0014208. The work at the University of Arkansas (quantum oscillation, topological physics, and surface state analyses) is supported by the US Department of Energy (DOE), Office of Science, Office of Basic Energy Sciences under Award DE-SC0019467. A portion of this work was performed at the National High Magnetic Field Laboratory, which is supported by the National Science Foundation Cooperative Agreement No. DMR-1157490, DMR-1644779, and the State of Florida.

**FIGURE CAPTIONS**

**FIG 1. ZrSiSe crystal structure and microscopy characterizations. a,** Crystal structure of ZrSiSe, showing the Se-Zr-Si-Zr-Se slabs and the cleavage plane (red arrow) **b,** Optical microscope image of a 28.2nm ZrSiSe nano-flake on Si/SiO$_2$ wafer obtained through micromechanical exfoliation. Inset, atomic force microscope image of a Hall bar device. **c, d,** Atomic resolution annular dark-field (ADF) aberration-corrected scanning transmission electron microscopy (STEM) images of (**c**) the bulk along the [100] zone and (**d**) exfoliated ZrSiSe flakes along the [110] zone. Inset in **c**, the [100] zone (cross-section) image matches well with the crystal structure.

**FIG 2. Thickness dependent SdH oscillations. a,** The oscillatory components $\Delta\rho_{xx}$ of thick (176nm) and thin (28.2nm) ZrSiSe samples with magnetic field normal to the sample surface. Inset, the fast Fourier transform of the corresponding oscillation patterns. The FFT for the 28.2nm sample is normalized to the 176nm sample according to the $F_B$ for clarity. The additional frequency of $F_S$=445T appears for the thin

sample. **b,** Thickness dependence of the relative FFT amplitude (FFTA) between $F_S$ and $F_B$ bands. Inset, corresponding FFT spectrums for different thicknesses, normalized to the 176nm sample according to the $F_B$ peak for clarity. **c,** Calculated Fermi surface cross-section at $k_z = 0$ of a three-layer ZrSiSe. The surface Fermi pocket is labeled in red. **d,** Calculated energy band dispersion of a three-layer slab ZrSiSe near X. The red color denotes the contribution from the surface state.

**FIG 3. Angular-dependent SdH oscillations. a,** Angular-dependent oscillatory components $\Delta\rho_{xy}$ of the 36nm ZrSiSe device at $T = 0.6$ K. Inset, the measurement setup. **b,** Fast Fourier transform of the SdH oscillation pattern shown in **a**. **c,** Polar plot of the angular-dependence of bulk frequency $F_B$ (blue) and surface frequency $F_S$ (red) from all measured devices (see Supplementary Fig. 4 and 5 for oscillation patterns and their FFT). The dashed straight line (red) indicates the $1/\cos\theta$ dependence for $F_S$.

**FIG 4. Property comparison between the bulk and surface bands. a,** Landau Level (LL) fan diagram for the bulk $F_B$ and surface $F_S$ states of four ZrSiSe nanodevices with thicknesses of 28.2nm, 33.9nm, 36nm, and 46.2nm. The solid lines represent linear fits of the Landau indices, which intercept around 0 for the surface band and -0.34 for the bulk band. Inset, zoom-in view showing different intercepts for surface and bulks. **b,** Berry phases derived from the LL fan diagram shown in **a** for samples with different thicknesses. **c,** The temperature-dependence of the FFT amplitude for bulk and surface bands for the same four samples in **a**. The solid lines indicate the fits to the thermal damping term of the LK-model. **d,** Effective masses for bulk and surface states derived from the fitting shown in **c**. **e,** Schematic drawing of the multichannel contributions to the Hall effect. Here, bulk e-, bulk h+, and surface e- denote contributions from bulk electron, bulk hole, and surface electron, respectively. **f,** Thickness dependent longitudinal conductivity and Hall coefficient. The solid lines show the fits to the three-channel model (see Methods).

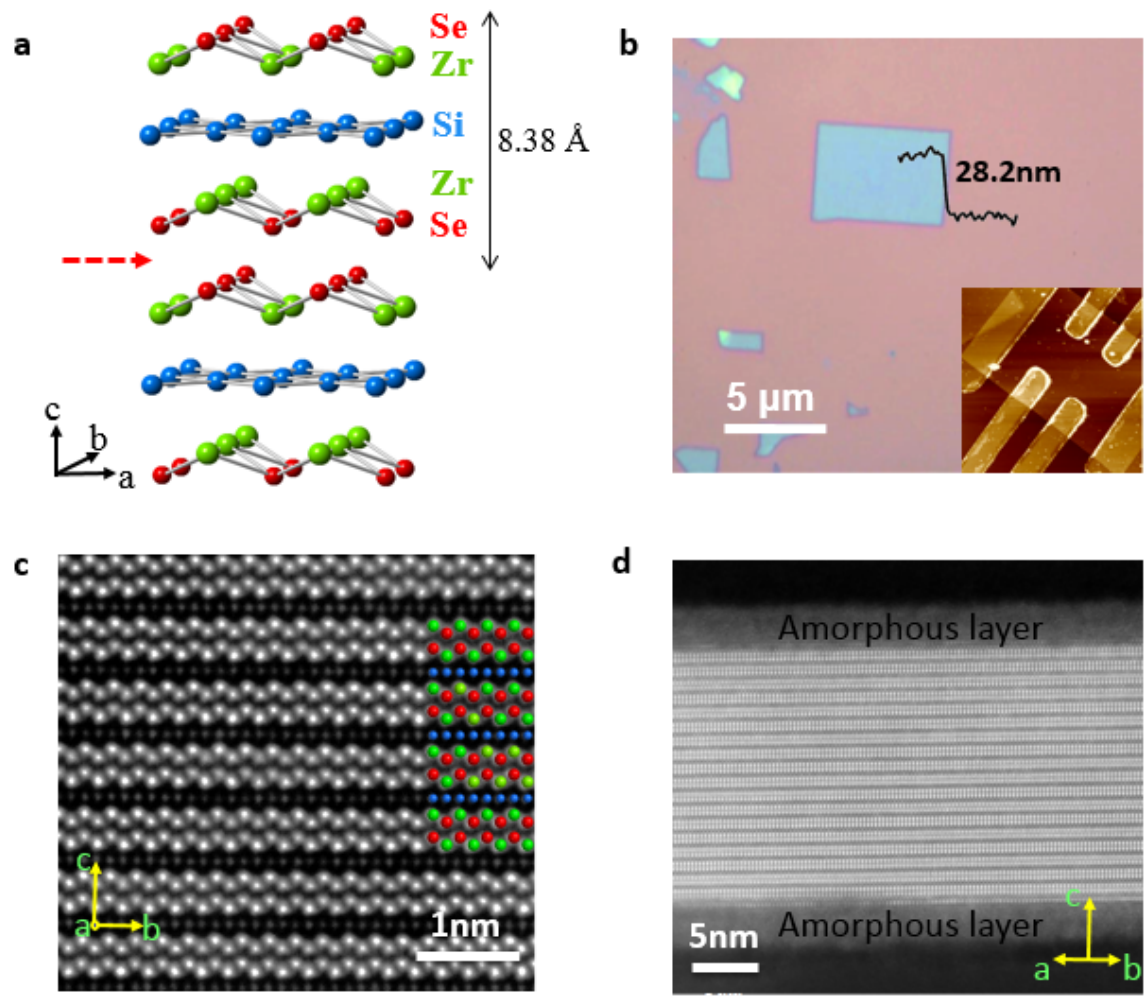

**Figure 1**

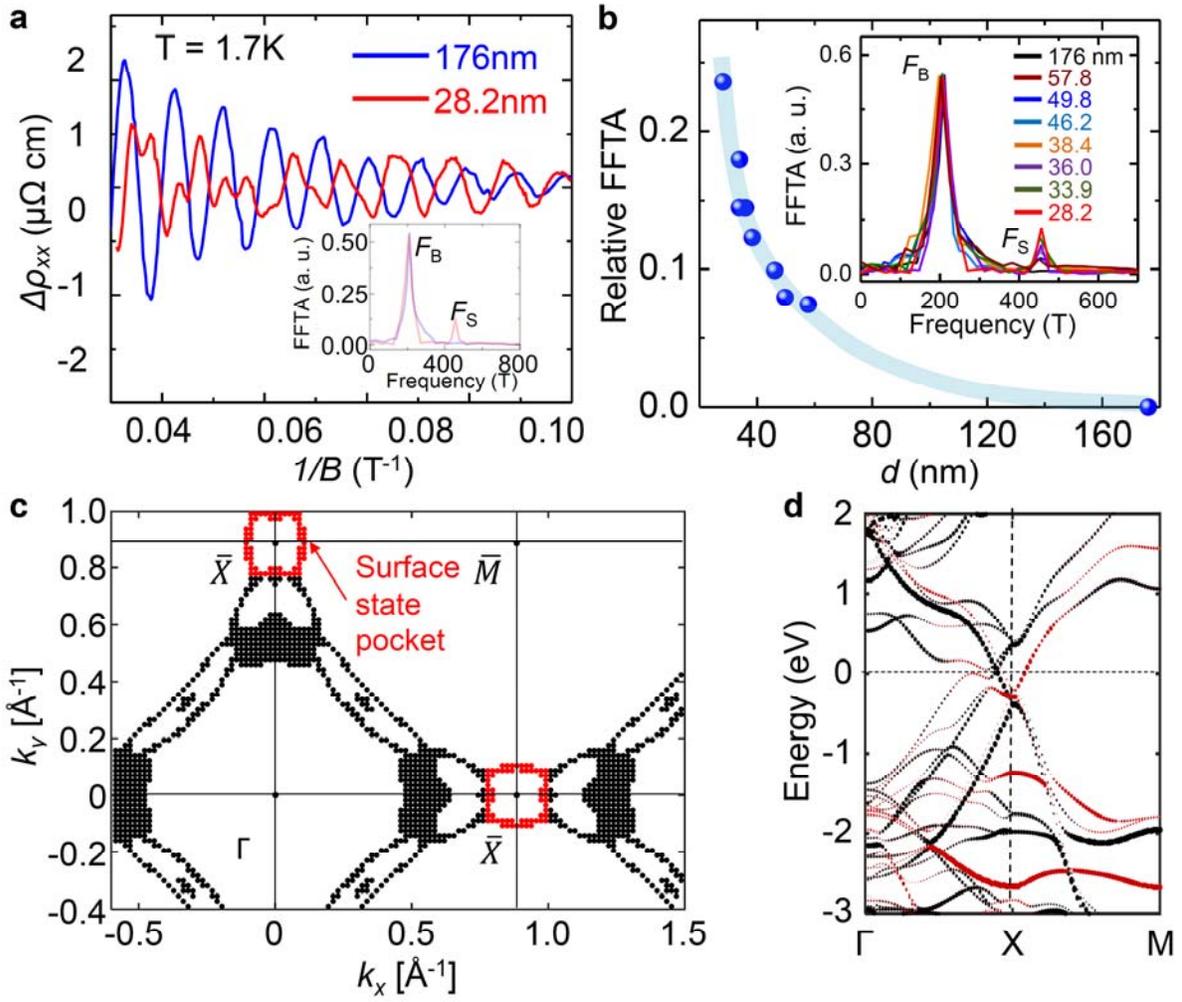

**Figure 2**

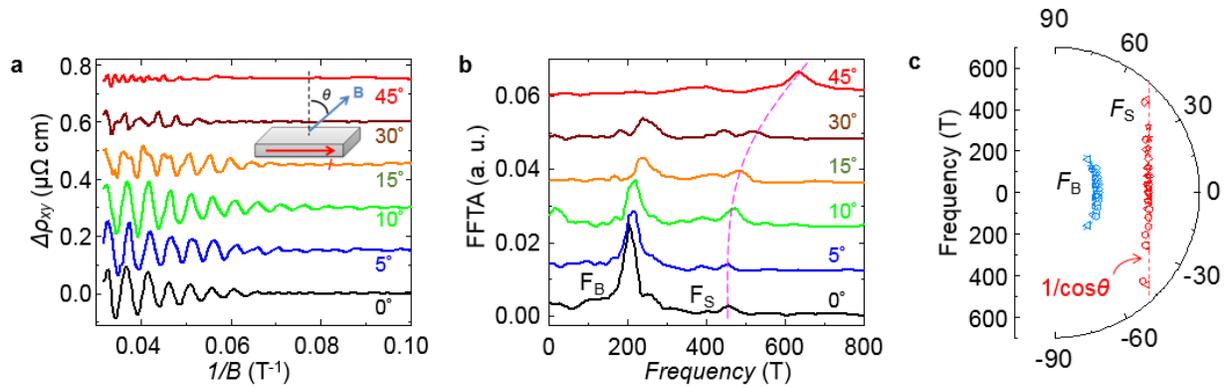

**Figure 3**

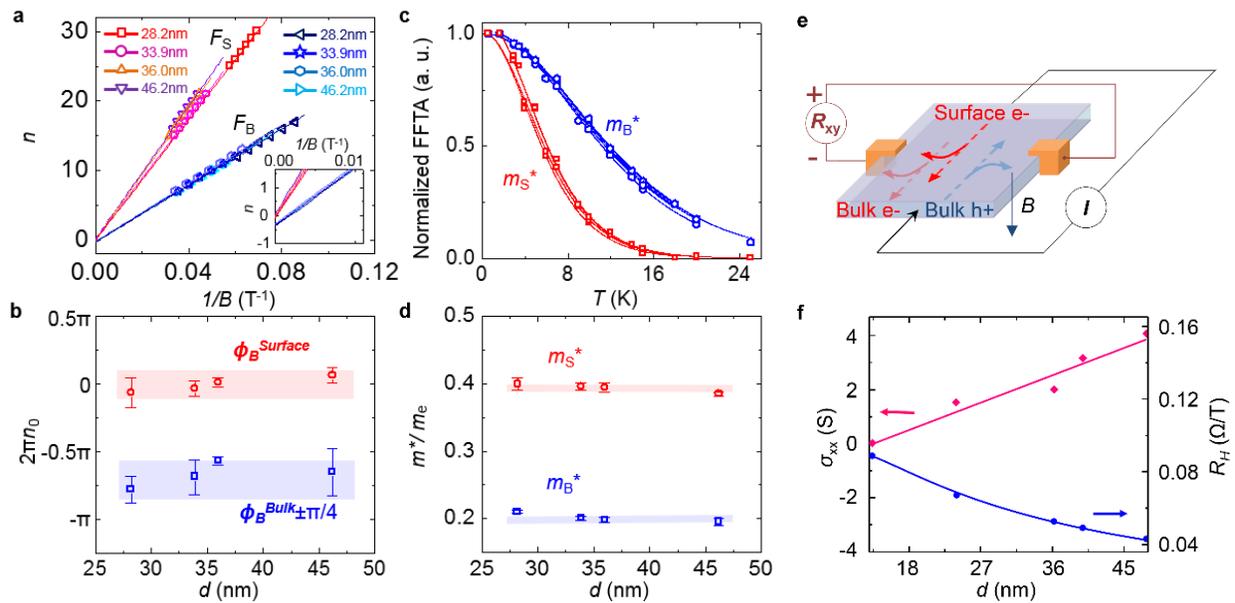

**Figure 4**